# The Mars Science Laboratory record of optical depth measurements via solar imaging

M.T. Lemmon [a] [*], S.D. Guzewich [b], J.M. Battalio [c], M.C. Malin [d], A. Vicente-Retortillo [e], M.-P. Zorzano [e], J. Martín-Torres [f] [g], R. Sullivan [h], J.N. Maki [i], M.D. Smith [b], J.F. Bell III [j]

[a] Space Science Institute, 4765 Walnut St, Suite B, Boulder, CO 80301, USA.

[b] NASA Goddard Space Flight Center, 8800 Greenbelt Rd, Greenbelt, MD 20771, USA.

[c] Department of Earth and Planetary Sciences, Yale University, 210 Whitney Ave. New Haven, CT 06511, USA.

[d] Malin Space Science Systems, Inc., P. O. Box 910148, San Diego, CA 92121, USA

[e] Centro de Astrobiología (CAB), CSIC-INTA, 28850 Torrejón de Ardoz, Madrid, Spain.

[f] Department of Planetary Sciences, School of Geosciences, University of Aberdeen, Aberdeen AB24 3UE, UK.

[g] Instituto Andaluz de Ciencias de la Tierra (CSIC-UGR), 18100 Armilla, Granada, Spain.

[h] Cornell University, Ithaca, NY, USA.

[i] Jet Propulsion Laboratory, California Institute of Technology, Pasadena, CA, USA

[j] Arizona State University, Tempe, AZ, USA

[*] Corresponding author, email address: MLemmon@SpaceScience.org

# Highlights

- The derivation of Curiosity's >5 Mars-year record of optical depth is described.
- The diurnal variation of the dust shows thermotidal effects during $L_S$ 180-360°.
- Water ice may contribute about 25% of the optical depth during $L_S$ 90-135°.
- The lack of scattering halos in ice hazes indicates small or amorphous particles.

# Keywords

Mars, atmosphere; Atmospheres, composition; Meteorology

# Abstract

The Mars Science Laboratory *Curiosity* rover has monitored the Martian environment in Gale crater since landing in 2012. This study reports the record of optical depth derived from visible and near-infrared images of the Sun. Aerosol optical depth, which is mostly due to dust but also includes ice, dominates the record, with gas optical depth too small to measure. The optical depth record includes the effects of regional dust storms and one planet-encircling dust event, showing the expected peaks during southern spring and summer and relatively lower and more stable optical depth in fall and winter. The measurements show that there is a seasonally varying diurnal change in dust load, with the optical depth peaking in the morning during southern spring and summer, correlated with thermotidal pressure changes. However, there was no systematic diurnal change during autumn and winter, except after one regional storm. There were indications that the dust was relatively enhanced at high altitudes during high-optical-depth periods and that high-altitude ice was significant during winter. The observations did not provide much information about particle size or composition, but they were consistent with a smaller particle size after aphelion (in southern winter). No scattering halos were seen in associated sky images, even when there was visual evidence of ice hazes or clouds, which suggests small or amorphous ice particles. Unexpectedly, the measurement campaign revealed that the cameras collected saltating sand in their sunshades 1.97 m above the surface. As a result, the measurement strategy had to be adjusted to avoid high-elevation imaging to avoid sand covering the optics.

# 1. Introduction

Since 2012, the Mars Science Laboratory (MSL) mission has monitored the local environment within Gale crater (5.4° S, 137.8° E) on Mars [Vasavada, 2022]. From the middle of Mars year (MY) 31 to late MY 36 (using the Clancy et al. [2000] convention), the *Curiosity* rover has traversed from its landing site on the plains of Aeolus Palus at the bottom of the crater and has ascended ~650 m of elevation up the slopes of Aeolus Mons, the 5-km mountain in the center of the crater. The ~29-km traverse was designed for a geological investigation, but a parallel atmospheric investigation has been characterizing weather at daily, seasonal, and inter-annual scales.

Among the key meteorological variables tracked by instrumentation on the rover was aerosol optical depth, which is primarily dust optical depth at the *Curiosity* location [Martínez et al., 2017]. Optical depth measurements at the surface are useful for providing ground truth for orbital measurements [Montabone et al., 2015] and for providing context for local measurements of other quantities. Such measurements tracked the planet-encircling dust event (PEDE) of MY 34 [Guzewich et al., 2019; Viúdez-Moreiras et al., 2019; Smith et al., 2019] and helped measure changes in the size of the dust aerosol [Vicente-Retortillo et al., 2017; Lemmon et al., 2019; Chen-Chen et al., 2021]. Precursor activity to dust storms has been studied using comparisons of the optical depth record and pressure fluctuations [Zurita-Zurita et al., 2022]. The optical depth record has enabled or enhanced studies of atmospheric tides [Harri et al., 2014; Guzewich et al., 2016], atmospheric composition [Webster et al., 2015; McConnochie et al., 2018], surface energy budget [Martínez et al., 2021], surface thermal inertia [Hamilton et al., 2014; Vasavada et al., 2017], dust deposition on surfaces [Vicente-Retortillo et al., 2018; Yingst et al., 2020], surface materials [Johnson et al., 2015; 2022], and the ultraviolet flux at the surface [Smith et al., 2016]. Comparisons of line-of-sight opacity with the column optical depth have shown increased dust loads have been concentrated at higher altitudes during regional storms, while dust lifting within the crater has made the crater a net source of dust aerosol at other times [Moore et al., 2019; Smith et al. 2020].

The purpose of this paper is to provide a documented public record of the time series of optical depth that MSL has provided and to explore some implications of the measurements. This paper covers imaging optical depth measurements during the MSL mission from landing through sol 3642, where a sol is a Martian solar day (roughly 24 hours and 40 minutes). First, the paper describes the observational dataset, including the images, their processing, and the *Curiosity* rover operations context that affected the investigation and its results. Then, it describes the calibration procedure. Finally, it discusses the results, their implications for seasonal and diurnal variability, the implications of observations at different wavelengths and low-elevation imaging, and the null result from a search for a water-ice scattering halo.

## 2. Observations

### 2.1. Instrument and observation overview

Solar imaging has been used to measure aerosol (primarily dust) optical depth for Mars surface missions such as the Viking Landers [Pollack et al., 1977; Colburn et al., 1989], Pathfinder Lander [Smith and Lemmon, 1999], Mars Exploration Rovers (MER) [Lemmon et al., 2004; 2015], Phoenix [Lemmon, 2010], and Perseverance [Lemmon et al., 2022b]. A solar-imaging capability was provided for the MSL mission by its Mast Cameras (Mastcams), 34- and 100-mm focal length color and multispectral imagers mounted on and aimed by the rover's Remote Sensing Mast (RSM) [Malin et al., 2017].

Each Mastcam has a filter wheel with a broad visible-light ("clear") filter and several narrow filters from 440-1013 nm. Filter 7 in each camera has a $10^{-5}$ neutral density (ND5) "solar filter" in addition to filters at (central wavelength ± half-width at half-maximum) 880±10 and 440±20 nm in the left (M-34) and right (M-100) cameras, respectively [Bell et al., 2017]. The detector has a Bayer pattern of red, green, and blue (RGB) microfilters, with a blue, a red, and two green pixels in each 2x2 sub-array. All pixels in the sub-array are nearly equally sensitive to 880-nm light. Only the blue element of the sub-array is especially sensitive to 440-nm light. During the MY-34 dust storm in 2018, non-solar filters at 447 and 867 nm were used for some observations [Lemmon et al., 2019].

The images resolve the Sun, thus allowing direct and diffuse light separation and a Beer-Lambert-Bougher law derivation of optical depth from the extinction of direct sunlight, as described in section 3. The field of view of the M-34 is about 15°, and that of the M-100 is about 5°. The instantaneous field of view (projected pixel size) of the M-34 is 220 µrad, and that of the M-100 is 74 µrad. The Sun's apparent diameter from Mars is typically about 0.33°, which is 27 (80) pixels for M-34 (M-100). Thus, there are typically about 560 pixels on the Sun for M-34 and 1250 blue pixels on the Sun for M-100 (Fig. 1).

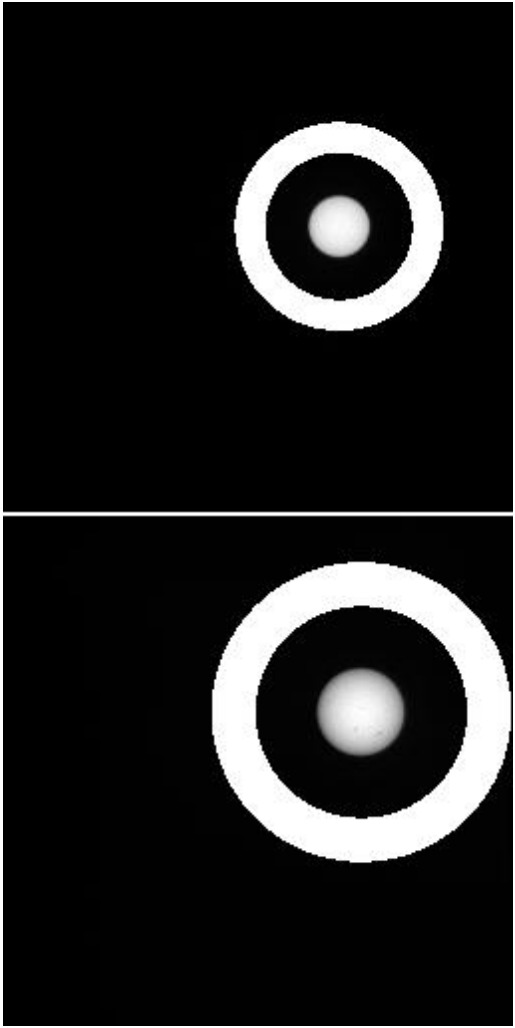

Figure 1. Sun images from M-34 (top) and M-100 (bottom) at 880- and 440-nm wavelength, respectively. The annulus outside the Sun shows the area used for background determination. The area interior to the annulus was used for flux extraction. There are sunspots visible in the M-100 image. The products used are 0226ML0010890030106050C00 (top) and 0226MR0010890020202883C00 (bottom).

## 2.2. Rover on-board processing and downlink

Mastcam image commands are divided into two main types: acquisition and transmission to the spacecraft. Image acquisition commands can generate autonomous transmission commands, and transmission commands can be sent separately. Unless otherwise specified, images are acquired losslessly but are compressed from 12-bit to 8-bit using a compression/expanding (companding) lookup table that has approximately a square root form of encoding (with the low data numbers handled more linearly). Sub-framing must be commanded at the time of image acquisition. That is, there exists no capability for an adaptive "Sun subframe" such as MER/Pancam had [Bell et al., 2003; Lemmon et al., 2015]. The resulting 8-bit image is grayscale. This grayscale image can be transmitted as a grayscale image, either losslessly or using monochromatic, lossy JPEG

(Joint Photographic Experts Group) compression. Alternatively, the 8-bit grayscale image can be expanded to 24-bit color (8-bit/band) and then compressed using color JPEG compression, with either 444 or 422 color encoding. In some cases, there may be multiple, complete image data products, such as when a lossy JPEG image is replaced by a re-transmitted, losslessly compressed image.

Thumbnail products contain 8x8 downsampled images, in which each 8x8 pixel JPEG compression block is replaced with their mean value for each Bayer color. Thumbnails were then usually compressed with color, lossy JPEG compression (e.g., a 1024x1024 pixel subframe would be represented as a 3-band,128x128 pixel thumbnail). Thumbnails of Sun images were returned to Earth quickly and were sometimes used for tactical optical depth estimates. In some cases, thumbnail products provided the ultimate optical depth measurement, as described below.

Once transmitted to Earth, all Mastcam images were converted to Experiment Data Records (EDRs, Malin [2013]) that represent the reconstruction of the detector array, and this paper uses the EDR archive through sol 3644. Near the end of a filename, an alphanumeric code indicates the data type and version number. The Alphabetical code includes "C" for losslessly compressed 8-bit data, "D" for lossy JPEG compressed grayscale, and "E" for color 24-bit JPEG compression. Versions increment from 01 to higher numbers based on the order of release, except that C00 is the highest quality, representing the lossless reconstruction of the array. Most 440-nm images after sol 990 were lossless, as were most 880-nm images after sol 2144. EDRs were used as the basis for optical depth processing to allow for flexibility in the calibration, as noted below.

### 2.3. Solar image data set

Solar imaging began on sol 33 at solar longitude ($L_S$) 168.5° in the late southern winter of MY 31 and is reported here through sol 3642 at $L_S$ 332.7° in the late southern summer of MY 36. The delay was due to the optical alignment of the Mastcams with the Chemistry and Camera (ChemCam), an instrument susceptible to damage from being aimed at the Sun under some circumstances. Procedures for ensuring the "Sun-safety" of ChemCam were not in place and adequately validated for operations before sol 33. While the MER project prioritized daily optical depth measurements to track solar panel performance [Lemmon et al., 2015; Lorenz et al., 2021], MSL had no such requirement. An approximately weekly cadence of observations was the initial goal. Early observations were typically paired within a sol to measure extinction with the Sun high and low in the sky, allowing a Beer-Lambert-Bouguer law determination of extinction. By around sol 1500, the strategy changed to acquiring observations on 2-3 sols per week. During 'dust storm campaigns,' observation frequency was increased to approximately daily.

Observations were typically scheduled to occur with other Mastcam imaging and varied in local time depending on the rover's schedule (Fig. 2). Early in the mission, observations significantly before noon local true solar time (LTST) were rare; eventually, alternating noon-afternoon and morning-noon pairs became more common, and ultimately a routine of one morning-block per week was established. However, the rover's location in a crater and varying local topography limited the earliest and latest local times. In addition, most sols had a period during late morning reserved for the rover to communicate with Earth and perform specific engineering tasks, which resulted in a lower frequency of observations during 9-11 LTST.

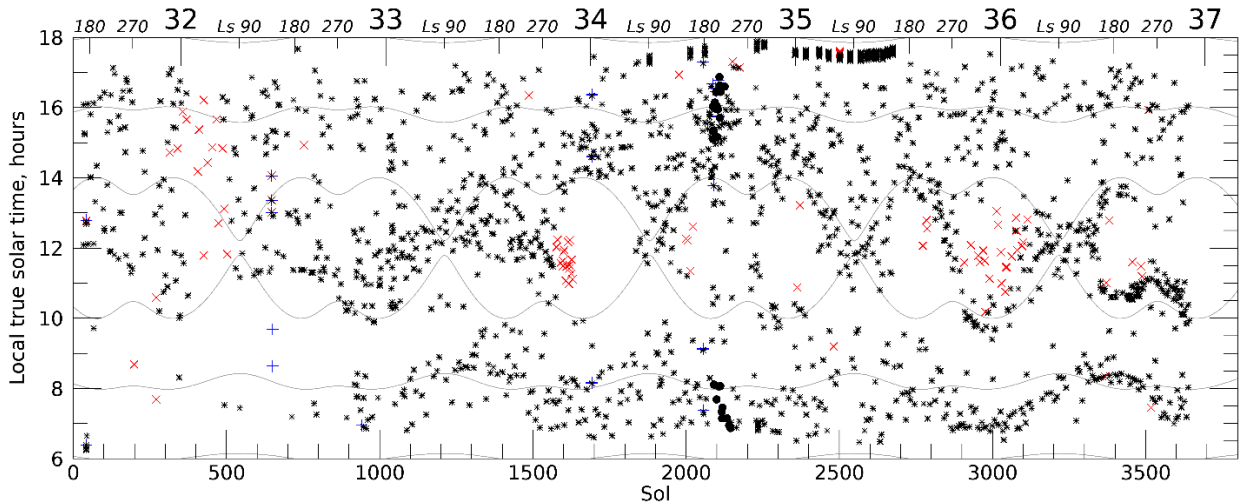

Figure 2. Optical depth sampling. Each observation's local time (LTST) is shown over the mission (Mars years are indicated along the top) as '×' for the 880-nm filter and '+' for the 440-nm filter. The symbols are overlaid in black for paired observations and appear singly in red or (rarely) blue, respectively, for unpaired observations. Circles indicate observations made with non-solar filters during sols 2086 to 2144. Gray contours indicate solar zenith angles of 30°, 60°, and 90°.

Observations were interrupted on a few occasions. After sol 198 (MY 31, $L_S$ 271°), a gap appears due to rover software issues precluding science observations. Holidays and later engineering issues, such as flight software upgrades, caused episodic interruptions. About every 26 months, there were gaps while Earth was near inferior conjunction with respect to Mars, and the Sun interfered with communications.

After sol 374 (MY 32, $L_S$ 12°), an approximation used on board the rover to determine the Sun's location had degraded to the point that the M-100 did not reliably have the Sun in its field of view, and images were suspended. After ~30 sols, M-34 imaging was resumed with a larger subframe and more compression; after ~100 sols, M-100 imaging was resumed with manual aiming. Finally, after another ~200 sols, a fix to the approximation was tested (involving some pairs of manual and automatically aimed Sun images), and regular Sun imaging resumed.

Between sols 1570 and 1634 (MY 33, $L_S$ 293-332°), while the rover was traversing the Bagnold dunes, sand entered and was trapped inside the cameras' baffles and began interfering with the measurements. The 440-nm optical depth seemed to increase by several tenths (and ultimately

>1.5) compared to the 880-nm optical depth, but only in mid-sol observations. An investigation determined that sand had entered the Mastcam baffles, which were designed as sunshades to minimize stray light in the cameras. Both baffles collected sand, despite their 1.97-m height. The M-34 baffle has a wide opening and presumably permitted more sand into the optics but could more easily shed sand when aimed down (as it typically was overnight). However, the smaller M-100 baffle aperture allowed less sand into the optics but was able to retain a higher volume of sand. As such, sand cascaded across the exterior optics when the cameras were aimed at very high elevation angles. Some maneuvers were attempted for sand removal, but they were largely ineffective. Continued sand obscuration was verified and monitored with a series of sky columns consisting of RGB images of the sky every 5° in elevation (see Supplementary Material). On sol 1634, the near-zenith transmission of the M-100 was 11% of nominal; this varied from 10% to 87% in 33 later measurements, with no pattern in location or season. Presumably, the detailed motions of the RSM, which were rover-tilt dependent for the sky column, partly controlled how much sand was deposited on the sapphire window, the outermost element in each optics. Efforts to clear the baffles were more successful with M-34: the sol-1634 near-zenith transmission was 80%, but obscuration was typically not measurable after sol 1700, with minor exceptions during southern summers.

After the sand issue was identified, M-100 images above 70-75° elevation were generally avoided, as were M-34 images above 80-85°. However, near-zenith pointing continued owing to Navcam sky imaging (e.g., Kloos et al. [2016]), so aeolian material trapped in the Mastcams continued to move across the camera windows. This may have happened with less sand before sol 1500; Mastcam has never had persistent spots from out-of-focus dust on the window like other cameras (cf., Chen-Chen et al. [2022]).

For the analyses described below, some images were screened out of the database due to issues such as high elevation angles with likely sand obscuration, low elevation angles that resulted in partial or complete obscuration by terrain, significant aiming errors that resulted in cropping or absence of the Sun, or a substantial number of saturated or non-linear pixels. In some cases, the images could have been used with modeling of the solar flux based on partial images, but this was judged to have low value due to the small number of images affected.

### 2.4. Accompanying sky images

Solar images were sometimes taken with accompanying sky images in the same or consecutive sequences. A 'basic tau' sequence acquired only solar images. A 'full tau' acquired solar images and paired or unpaired sky images in three sky positions. Sky surveys acquired a 'full tau' and images across the sky [Lemmon et al., 2019]. Early sky images were taken with 447- and 867-nm filters. Due to stray light issues in the 867-nm wavelength images (the 'black' interior of the baffle was reflective in the infrared), they were replaced by RGB images on sol 2452. While Lemmon et al. [2019] used 447-nm wavelength sky images to model particle size, that is beyond the scope of this work. However, the sky images will be used for context when appropriate.

# 3. Optical depth calibration

## 3.1. Overview

Our optical depth derivation followed the general pattern of Pollack et al. [1977] and every similar investigation since. It was explicitly adapted from the method described in detail by Lemmon et al. [2015]. Sun images provided radiance of the Sun with the sky and any uncorrected instrument background adjacent to the Sun. After background subtraction, integration over an aperture centered on the Sun determined the direct solar flux. Multiple images on a sol allowed a Beer-Lambert-Bougher law flux calibration; the flux calibration was averaged and tracked over time and used to reduce individual Sun images to optical depth. The discussion in the remainder of this section focuses on the areas where Mastcam differs from the MER Pancams, including the Bayer pattern and filter combination, the lack of a known temperature dependence from laboratory calibration, and the different history of exposure to the elements.

## 3.2. Image processing

The first step in determining optical depth was measuring direct solar flux in units that could be compared across images. Decompressed images were restored from 8-bit to 12-bit digital numbers (DN) based on an inverse lookup table. Estimated bias and dark current were subtracted, hot pixels and cosmic ray strikes were replaced with local-median values, and exposure time and the flat field were divided out before extracting the background-removed solar flux. Figure 1 illustrates the annulus used for background determination, with the area interior to the annulus used for aperture photometry. The annulus extends from 2.4 to 3.4 apparent solar radii from the center of the Sun. The solar radius (in pixels) used in processing scales inversely with the Sun-Mars distance and varies with image or thumbnail resolution. The background signal comprised residual instrument electronic bias, dark current, internally scattered sunlight, and occasionally detectable skylight. The background was typically ~0.5% of the peak signal. After the background subtraction, the sum of the signal interior to the annulus was taken, resulting in flux in cumulative DN/millisecond units, which was scaled to a reference Sun-Mars distance of 1.5 AU.

M-100 images posed unique challenges, due to the blue filter and light leaks. Lossless 440-nm images that return the detector readout with no color interpolation ('de-Bayering') were used to investigate using less sensitive Bayer green pixels. This failed to improve signal quality and caused an inconsistent comparison with thumbnail and other lossy images. As such, M-100 images were resampled using only the blue pixels (ignoring red and green pixels for lossless data and interpolated pixels for JPEG data), resulting in one-half the linear resolution. Also, for these images, light leaks through the ND5 filter affected some portions of the images. The targeting typically kept the Sun away from these areas; the areas were masked out in processing to avoid false detections of the Sun.

For thumbnails, the blue channel was used for 440-nm images, while the sum of the red, blue, and twice the green channel was used for 880-nm images (reflecting the Bayer unit cell's sampling). The 8-pixel square compression block in thumbnail images is larger than the apparent radius of the Sun, meaning that the compression could impact the photometry. For this reason, a high compression quality was used (JPEG quality 95). Even so, the photometry of the thumbnail images differed from that of the regular images systematically after scaling for the number of pixels (i.e., 64 image pixels map to one thumbnail pixel at 880 nm). For all image-thumbnail pairs, the thumbnail-to-image flux ratio was calculated. That ratio was, for each filter, found to be a noisy constant with no dependence on signal level, temperature, season, etc. For the 880-nm filter, the ratio was 0.9026±0.0006, with the standard deviation for individual measurements being 0.0118. For the 440-nm filter, the ratio was 0.9357±0.0007, with the standard deviation for individual measurements being 0.0131. Thus, thumbnail-image-derived fluxes were substituted for missing regular images after dividing by the ratio, with increased uncertainty.

The dominant uncertainty sources in the flux measurement were the thumbnail scaling, the lookup table's quantization of the solar and background radiance, the flat field correction that allowed the comparison of measurements taken from different pixels on the detector, and the effect of temperature. Thumbnail scaling added 1.3% uncertainty. The quantization terms typically amounted to 0.2% but could be several percent for under-exposed images. The flat-field term is likely around 1%, and the temperature term is no more than a few percent, as discussed below. While sunspots were observed in images, their effects on 440- and 880-nm fluxes were minor compared to uncertainties. The uncertainty from the radiometric conversion of DN/ms to physical flux units is discussed below.

### 3.3. Determining atmospheric transmission and optical depth

The instantaneous optical depth along the line of sight from the camera to the Sun is,

$\tau_{path} = \ln(F_0/F_{observed})$,

where $F_0$ and $F_{observed}$ are the solar flux—at the top of the atmosphere and observed—in the same units. Normal optical depth is path optical depth divided by the airmass. While the secant of the zenith angle is a good approximation for airmass generally, to deal with low-elevation images, the airmass was initially computed for a spherical-shell atmosphere with a scale height of 10.8 km for (colder) morning observations and 11.8 km for (warmer) afternoon observations based on the Mars Climate Database [Forget et al., 1999; Millour et al., 2015]. There is evidence for a more complex vertical structure [Moores et al., 2015; Moore et al., 2016; Smith et al., 2020]; however, nearly all the observations would be insensitive to the distinction. For the subset of low-elevation observations sensitive to the vertical structure, an effective scale height of 13 km was ultimately derived as described below.

The simplicity of determining $F_0$ and thereby retrieving optical depth was complicated by several factors. 1) The rate of DN accumulation from a fixed source should vary with temperature, but this was not measured pre-flight. Bell et al. [2003] showed a linear relationship of the response function with temperature for Pancam, as did Hayes et al. [2021] for Mastcam-Z, which used the same detectors as Mastcam. Heaters generally limited the Mastcam temperatures to >-20°C (>-30°C after sol ~2500); the low, mean, and high observed temperatures were -39, -17, and 11°C. 2) Performing a Beer-Lambert-Bouguer law calibration on Mars relies on an assumption of non-varying optical depth during some period. However, Haberle et al. [2014] showed a typical 10% pressure change over a sol. This would imply a 10% dust load change for well-mixed dust or optical depth changes of more or less than 10% if more or less dusty air contributed to the changes. 3) The Mastcams, while partly protected by baffles, interact with sand and dust in their environment. This could cause a clear-sky measurement to vary over time and look like a varying $F_0$. Lemmon et al. [2019] reported that the 440-nm to 880-nm optical depth ratio appeared to change around sols 1450-1600, around the time when sand was found in the baffles.

To account for the temperature sensitivity of the detector, we relied on analog values interpolated from the Hayes et al. [2019] calibration of Mastcam-Z. Early calculations made this a parameter to be fit for each filter.

Complexity in the calibration was added by the likelihood of a systematic variation of optical depth with pressure and, therefore, time of sol. If optical depth varied randomly or only by small amounts (compared to desired accuracy), then a series of linear fits of $\ln(F_{observed})$ versus airmass would provide a series of fit $F_0$ from the intercepts. However, given the relative abundance of morning and afternoon data, the intercept would be wrong if the optical depth at 17:00 LTST were always 5% less than at 12:00 LTST. As shown by Haberle et al. [2014] and inspection of the pressure record [Gomez Elvira, 2013], the pressure falls almost linearly with LTST over 08:00-16:00, albeit with different slopes at different seasons due to varying pressure tides [Guzewich et al. 2016]. To derive a flux calibration, we use an effective $\tau_{noon}$ and assume that optical depth varies linearly during the sol. We stress that this is only a tool to combat a possible source of systematic error. Actual optical depth values were computed with no time-of-sol correction.

Several models for the behavior of $F_0$ through the mission were considered. Based solely on the set of individual fits, no variation was required. Adding free parameters (such as the piecewise-linear fit necessary for Pancam processing in Lemmon et al. [2015]) improved the fit only moderately, with no statistical significance. However, we have seen that the optical depth ratio changed when a new environmental impact on the detector was observed. Therefore, based on inspection of initial calculations of the optical-depth ratio, we used one calibration parameter before sol 1450, one after sol 1550 until sol 3642, and a linear change in between.

To evaluate calibration parameters, every measurement was reduced to an implied noon-normal optical depth based on $F_0$, the slope of response with temperature, and the slope of dust load with hour:

$$\tau_{noon} = -\ln\left(\frac{F_{observed} \cdot (1 + P_1 T)}{F_0(S)}\right)/\eta + P_2(LTST - 12)$$

where $\tau_{noon}$ is the estimated reference optical depth at noon; $T$ is the temperature in °C; $S$ is the sol number; $\eta$ is airmass; $LTST$ is the local true solar time, limited between 8 and 16; and the $P_1$, $P_2$, and two $F_0$ values representing early and later behavior are the calibration constants. Omitting the $P_2$ term results in the instantaneous optical depth, $\tau_{observed}$.

Parameters were tested as follows. For every 3-sol period in which there were multiple measurements but no rapid changes due to local storms were observed, errors were determined as the residual of noon optical depth compared with the period's mean value. Instrumental uncertainties and an additional uncertainty representing a 5% natural variation in optical depth [Lemmon et al., 2015] were included in the error budget. Parameters that minimized $\chi^2$ are shown in Table 1. The resulting optical depths are shown in Fig. 3 (for the data archive, see Lemmon [2023]).

Table 1. Calibration constants. $P_1$ was fixed based on Hayes et al. [2019]. Note that $P_2$ is an intermediate parameter provided for reference only and was not used to determine reported optical depth.

| Constant | M-100 | M-34 |
| --- | --- | --- |
| Filter, wavelength | R7, 440 nm | L7, 880 nm |
| $F_{0,1.5\ AU}$, sol<1450, DN/ms | (3.171±0.111) x$10^5$ | (2.827±0.099) x$10^4$ |
| $F_{0,1.5\ AU}$, sol>1550, DN/ms | (3.064±0.107) x$10^5$ | (2.906±0.102) x$10^4$ |
| $P_1$, °$C^{-1}$ | 0.0010±0.0006 | -0.0030±0.0006 |
| $P_2$, hour$^{-1}$ | 0.0135 | 0.0116 |

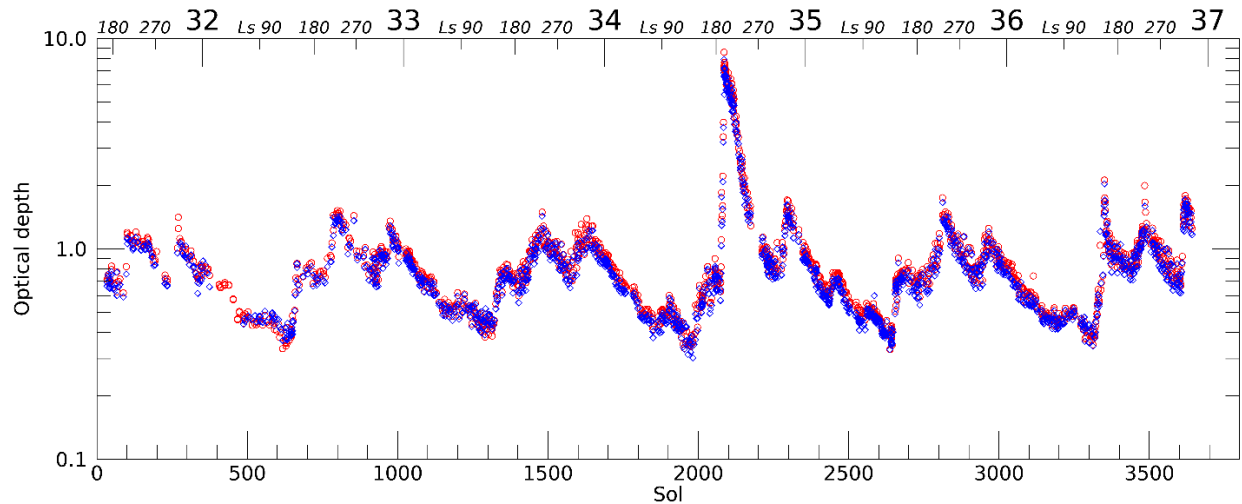

Figure 3. The optical depth over the mission (Mars years are indicated along the top) is shown for 440 nm (blue diamonds) and 880 nm (red circles).

## 3.4. Extension to non-solar filters in dusty conditions

During the 2018 dust storm, high optical depths were reached over an extended period. This required modifying the standard protocol to acquire images away from noon: for example, at airmass three and normal optical depth four, the atmospheric extinction was larger than that of the ND5 solar filters. Nonetheless, the direct flux could be measured independently of the diffuse sky flux with a solar image. Therefore, we developed operational guidelines that the solar filters could be used up for path optical depths <11.5, while some other filters could be used for path optical depths of 9.6 to 18 (above which the Sun could not be seen). As a result, eight long-exposure images in the solar filters were paired with short-exposure images in the L5 and R2 filters at 867 and 447 nm, respectively. These were used to cross-calibrate the filters and infer from the non-solar filters what flux the solar filters would have seen. This process ignored the slight wavelength difference between solar and non-solar filters, but Fig. 3 shows that the 440- to 880-nm wavelength variation is small (also see section 4.3 for wavelength variations). The L5:L7 flux ratio was determined to be $(4.75\pm0.07) \times 10^4$, and the R2:R7 flux ratio was determined to be $(1.92\pm0.04) \times 10^5$.

## 3.5. Uncertainty

Uncertainty was assigned to each value based on uncertainties in the fit parameters and measurement characteristics. Path optical depth uncertainty was based on the uncertainties in the flux measurement (usually <1%), $F_0$, and temperature correction. The uncertainty in $F_0$ was taken to be 3.5% based on the variance in results of a Beer-Lambert-Bouguer-Law fitting to the temperature- and time-of-sol-corrected data. The uncertainty in the temperature correction was taken to be 0.0006 $°C^{-1}$ based on tests with freely varying temperature corrections. For the conversion to normal optical depth, airmass uncertainty was computed. For this, we computed the airmass conversion for scale heights of 10 and 16 km for each solar elevation and associated that with a 2-sigma error; the choice of scale heights is discussed in section 4.4.

Given that time, temperature, and airmass were partly correlated, a further investigation of uncertainty was conducted. A preliminary determination of sol-average optical depth was taken as a reference (or 'truth') condition. Four hundred simulated measurement sets were constructed using these sol-average optical depths with random variations and diurnal perturbations added. The resulting flux was randomly varied based on the flux uncertainty and perturbed based on temperature and sol. Randomly varied calibration parameters determined each simulation. For each simulated measurement set, calibration parameters and optical depths were derived. These were compared to the input calibration parameters and truth optical depths. We found that bias in each parameter (defined as the difference between the reference parameter and the mean retrieved parameter) was 1-3 orders of magnitude smaller than the standard deviation of the retrieved parameter. The standard deviation of each retrieved parameter was consistent with the stated uncertainties, and the standard deviation of the retrieved optical depths was consistent with their calculated uncertainties.

# 4. Results and discussion

## 4.1. Seasonal variations of optical depth

Figure 4 shows the 880-nm optical depth record as a function of the season for each MY of observations. Given how closely they track the 880-nm measurements, the 440-nm measurements were omitted; the wavelength dependence is discussed in 4.3. High optical depths in the 2018 dust storm are truncated but were discussed in detail by Guzewich et al. [2019] and Lemmon et al. [2019]. The observed values show expected features from other landed and orbital assets [Kahre et al., 2017]. Optical depths declined to low values during $L_S$ 0-140°. Dust generally increased around $L_S$ 140-150°, and dust increased around regional storm activity near $L_S$ 210° and 315°. The increase around sol 664 (MY 32, $L_S$ 148°) was associated with dust storms to the south and a convective vortex outbreak at the rover's location [Kahanpää et al., 2017].

Morning and afternoon measurements are shown as open and filled symbols, respectively, in Fig. 4. Over $L_S$ 0-150° (and 150-180° except for MY 36), the morning and afternoon trends were not easily distinguished. However, over most of the rest of the year, morning values had a trend above that of the afternoon values, in particular during the declining phase of dust storms, such as the MY36 $L_S$-150° storm, the MY34 $L_S$-180° storm, and the MY34 $L_S$-320° storm. Diurnal effects are discussed further in section 4.2.

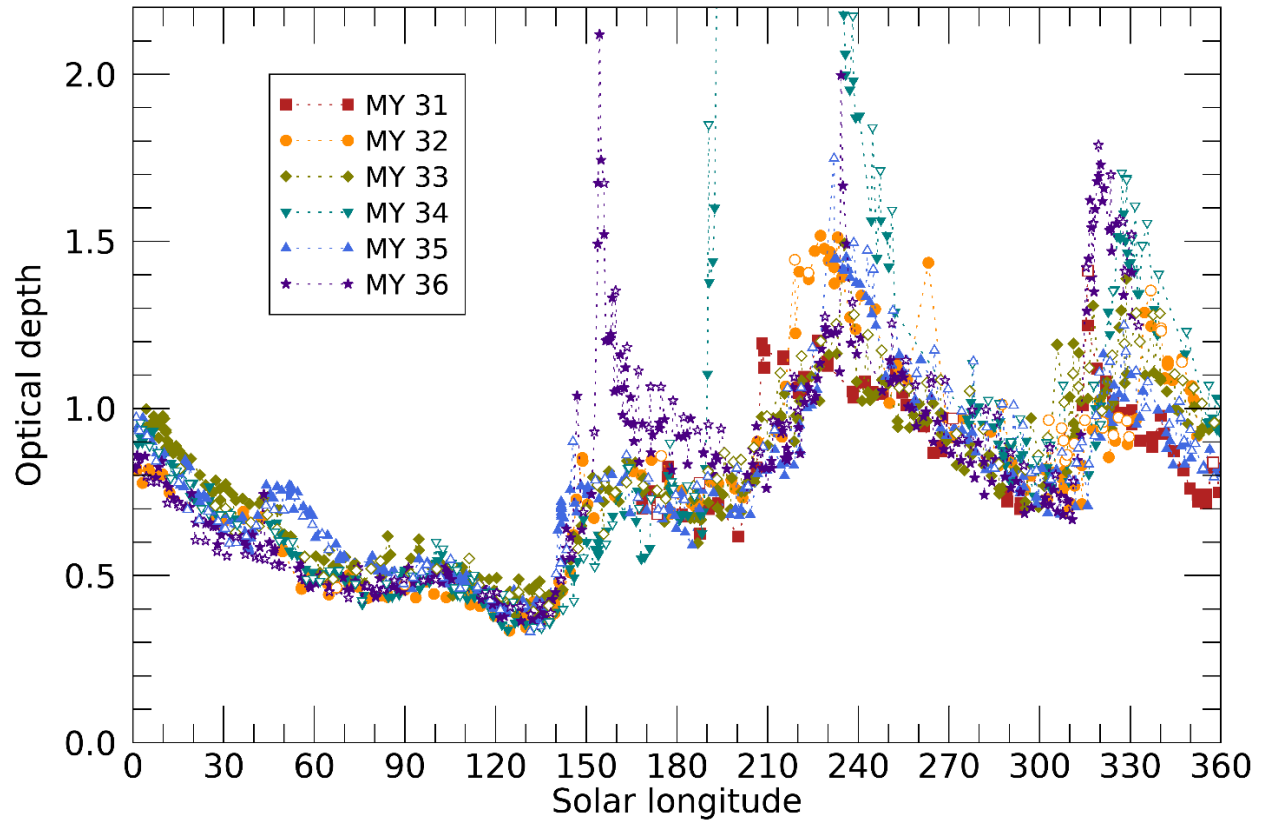

Figure 4. The 880-nm optical depth is shown for each Mars Year (MY) of the mission. Open symbols show data from before 10:00 LTST, and filled symbols show all other data.

Beyond the PEDE of MY34, the only unusual storm was the $L_S$-150° dust storm of MY36. This was primarily experienced as increased dustiness from elsewhere, without the active lifting seen by Perseverance [Lemmon et al., 2022b]. *Spirit* saw a similar storm, but that time frame does not usually have such high optical depths, instead usually showing a moderate optical depth increase. Around the $L_S$ 210° and 315° storm seasons, there was variability in the onset and peak for each period, but each year looks qualitatively like the others.

There are several instances during $L_S$ 40-140° in which an isolated measurement was ~0.1-0.2 above nearby measurements. These were likely ice clouds or hazes along the line of sight to the Sun. That period, especially, may also include an ice haze rather than solely dust, but the measurements do not distinguish the source of optical depth. However, the optical depth ratio in the two filters is potentially sensitive to a change in composition; section 4.3 describes a systematic difference over $L_S$ 90-140° that may be due to ice. However, even this is speculative.

### 4.2. Diurnal variations of optical depth

An inspection of Fig.4 suggests that there is a diurnal variation during parts of the year. While the result of the flux calibration suggested that a diurnal variation would be found, the calibration does not impose any such variation. One top-of-atmosphere flux is applied regardless of the time of sol.

We divided the data into bins of 30° of L$_S$ to investigate the diurnal variability and its seasonal behavior. For each bin, we fit the optical depths to a cubic polynomial to divide out secular trends after masking out data associated with sudden changes (storm onsets in MY34, 36). The residuals are shown in Fig. 5 for each period including at least ten measurements to support the detrending. A linear fit in LTST is also shown as a solid line. Generally, the end of MY31 had relatively flat trends but with relatively little morning data. For subsequent data, the first half of each MY had somewhat flatter slopes, and the second half had steeper slopes, showing an optical depth decline during the sol, similar to the pressure trend. Figure 6 shows the distribution of slopes in each half of an MY: the early part of the year had a modal slope of zero; the second half had a modal slope of -1.5%/hour. The distributions were narrow compared to the difference.

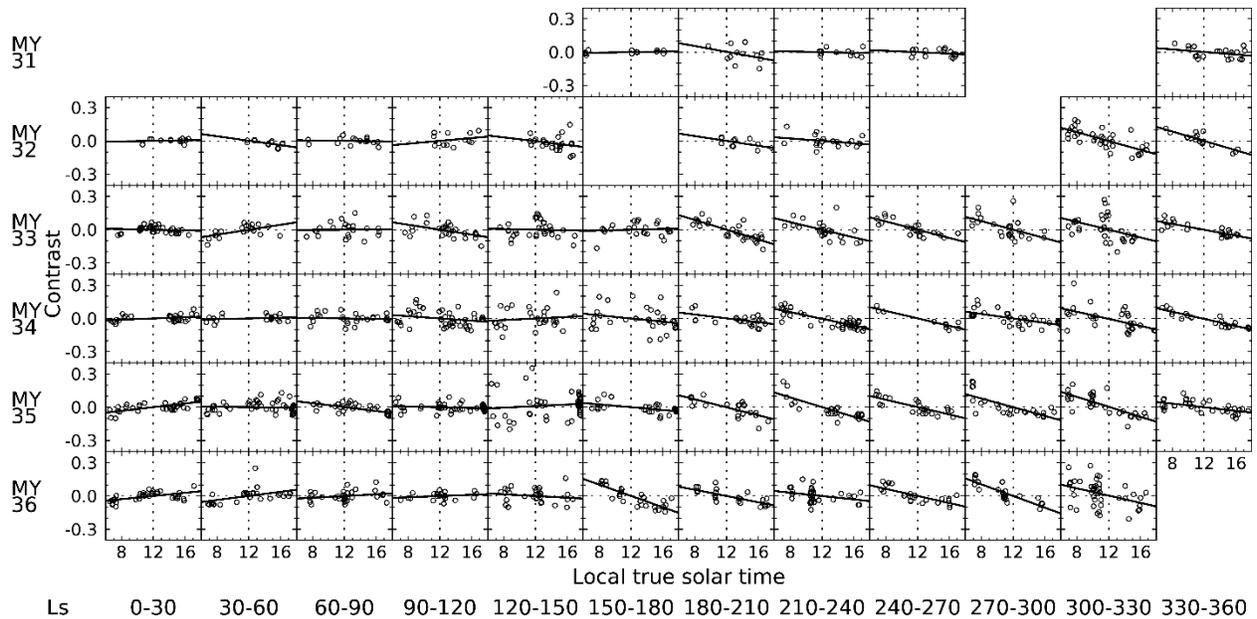

Figure 5. Detrended optical depths are shown in 30°-L$_S$ intervals as a function of LTST, with a linear fit shown.

There is a pressure slope with LTST due to tides, with a mean drop during the first 100 sols of the mission of 85 Pa over 08:00 to 16:00 LTST, representing a range of 11% of the mean pressure of 775 Pa [Haberle et al., 2014]. While the pressure change is larger in the year's second half, it is always present [Guzewich et al., 2016], and the tides are amplified within the crater [Richardson and Newman, 2018]. During the second half of the Mars year, the observed optical depth slope was consistent with an approximately constant dust mixing ratio for each sol. The optical depth decreased by ~12% from 08:00 to 16:00, as would be expected if the dust were moving with the air moving in response to thermotidal variations. However, only an influx of relatively dust-free air (presumably above the planetary boundary layer) could increase the pressure without increasing the optical depth, as was observed during the first half of each MY. The difference between the two parts of the year seems to be that dust is well-mixed to higher altitudes through the year's second half when the planetary boundary layer (PBL) grows to exceed the crater's depth [Fonseca et al., 2018]. In contrast, we speculate that air with lower dust content moves over the crater during most of the year's first half when PBL heights are minimal, such that the optical depth changes little despite diurnal pressure changes. The only exceptional slope within the year's first half was after the $L_S$-150° dust storm of MY 36.

No evidence of systematic diurnal variations of ice hazes during the first half of the Mars year was seen. While outlier points may indicate cloudy or hazy sols, there are not enough such points to cause an observable morning-afternoon difference. The expected variability is small compared to the uncertainty (typically ~8% of optical depth around $L_S$ 60-90°) and the natural variability seen in Fig. 5. Kloos et al. (2018) found a morning enhancement in average cloud optical depth of a few hundredths (<10% of total optical depth), with variability of several hundredths. Smith et al. (2023) found variations in infrared optical depths as high as 0.2-0.3 around $L_S$ 100° at Jezero crater, corresponding to visible optical depths of 0.3-0.5 using their multiplier of 1.61. This would represent a 60-100% enhancement in the morning, which is clearly ruled out by our data for the Gale crater site.

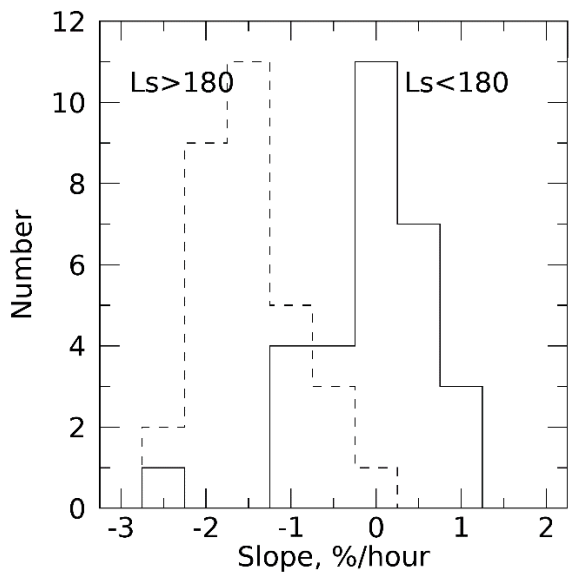

Figure 6. Histograms of the distribution of slopes from Fig. 5 are shown for $L_S$<180° (solid line) and $L_S$>180° (dashed line).

### 4.3. Implications of the wavelength dependence of extinction

While optical depth ratios can be sensitive to and sometimes diagnostic of particle size, visible and near-infrared optical depth ratios do not typically vary much for Mars's dust [Lemmon et al., 2015; 2019]. To test the variability, we converted measured optical depths to aerosol optical depths by removing Rayleigh scattering. Rayleigh optical depths in Mars's atmosphere are small

(almost 0.01 at 440 nm and <0.001 at 880 nm) compared to the measurement uncertainties but could be a source of systematic error if unaccounted for. The result (Fig. 7) shows the same pattern that Lemmon et al. [2019] noted, with an Angstrom exponent of -0.09±0.01 during most of the year. However, over $L_S$ 90-130°, the exponent increases to near zero, although the data are noisy during this period of low optical depths. This could be related to the sedimentation of large particles resulting in smaller particles in the atmosphere at that time [Vicente-Retortillo et al., 2017; Lemmon et al., 2019; Chen-Chen et al., 2021] or to ice clouds. However, the small and uncertain change is insufficient to make any determination about Martian aerosols, and future studies that intend to address this issue should use a larger wavelength baseline; otherwise, a single wavelength may suffice. We observed no morning-afternoon difference, and midsol data were of insufficient quality for any comparison due to low path lengths through the dust and ice.

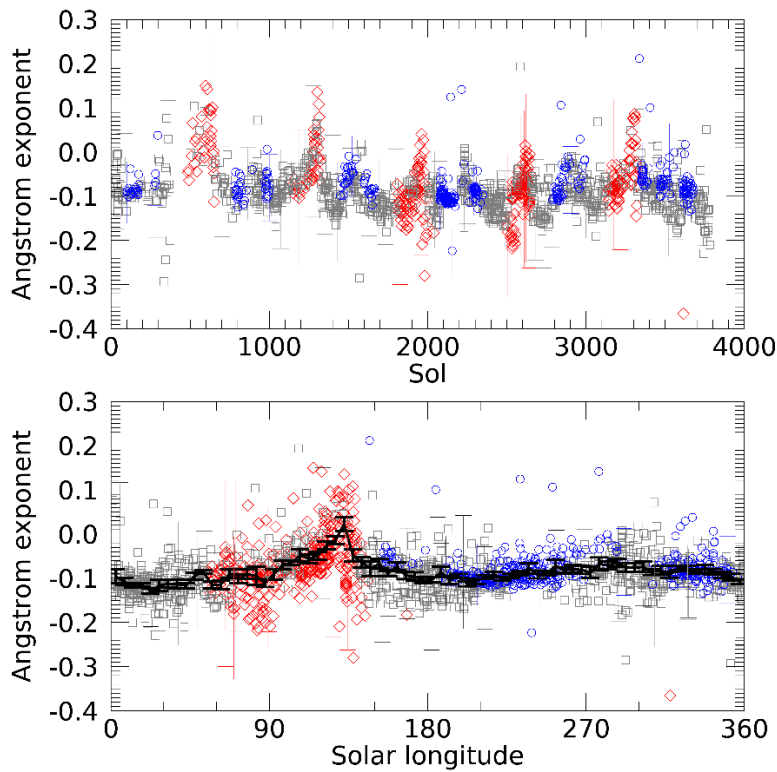

Figure 7. Angstrom exponent for 440- vs. 880-nm optical depth is shown as a function of sol (top) and $L_S$ (bottom). Measurements are color-coded by uncertainty in the optical depth ratio: >20% (red), <5% (blue), and other (gray). Error bars are shown only for every 20th point. The thick black line on the bottom shows the data binned every 5° in $L_S$.

### 4.4. Implications of high-incidence angle observations

Low-Sun observations are sensitive to the vertical structure of the atmosphere in that the path optical depth corresponding to a particular normal optical depth decreases with increasing scale height. For Mars, this effect is minor above 20° elevation angle, but the vertical structure becomes the dominant source of uncertainty below 11° in this data set. Since uncertainties in path optical depths tend to be similar across measurements, uncertainty in normal optical depth was minimized at 15°. Above this, the uncertainty is divided by a smaller airmass, while below this, the airmass dominates the uncertainty.

Several observations were designed to test the assumption that the dust was well mixed, at least in the first few scale heights—the observations are not sensitive to a declining scale height in the upper atmosphere. These observations comprised a series of 7 image pairs designed to end just before the Sun hit the local horizon (and some included images of the Sun partly or wholly obscured and were screened from the database). These were run effectively only in the evening, as the presence of Aeolis Mons precluded low-elevation imaging in mornings by the time this was done in MY 35-36. Figure 8 shows the results of one such sequence on sol 2255. The data were fit with a scale height of 13.9±1.3 km; curves for 10 and 18 km are also shown. Figure 9 shows the fits to all sets of evening data with the Sun under 12° (low-quality fits were omitted, including occasions where the optical depth changed during the sequence based on inspection).

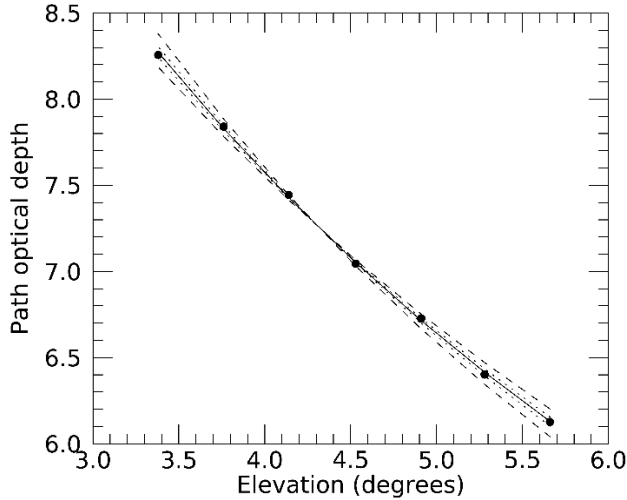

Figure 8. Path optical depth (product of derived optical depth and nominal airmass) is shown as a function of solar elevation angle for data from sol 2255 (circles, with uncertainties smaller than the symbol) and five models. The solid line shows a fit using H=13.9 km; dotted lines are 1-sigma (1.3 km) from that, and dashed lines are H=10 and 18 km; lower scale heights result in steeper slopes.

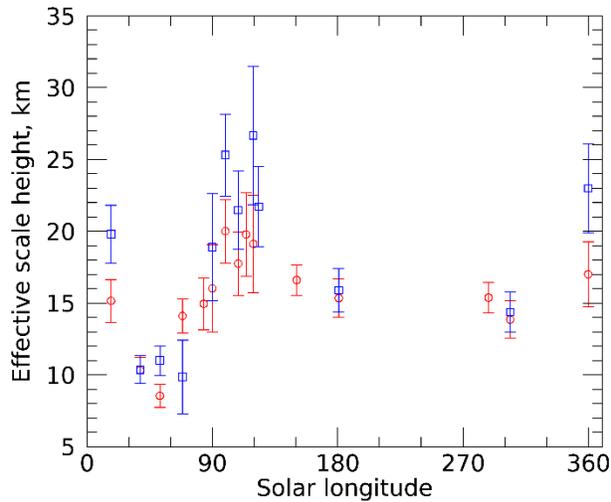

Figure 9. Best-fit scale heights for all evenings with >3 observations with the Sun under 12° are shown for the 440-nm (blue) and 880-nm (red) filters.

There are two types of implications of this analysis. One is for the data set. It became clear that the initial 10- to 11-km scale heights were unsuitable for obtaining comparable data from low- and high-Sun cases. So, the data were reprocessed with a mid-range value of 13 km, and the uncertainty was defined by calculations for 10- and 17-km scale heights. Nonetheless, data from below 10° should be used with caution. The second implication is for Mars's aerosols, which were at higher altitudes in some parts of the year than others.

We have not presumed the aerosols are distributed with the stated scale heights, but the scale-height changes were markers for vertical structure changes. Lemmon et al. [2015] noted that the scale-height model is a simple, 2-free-parameter model to describe the vertical distribution of dust. The data cannot constrain a model with well-mixed dust within the crater, a different mixing ratio above the crater, and other dust or ice-haze layers at higher altitudes. So, the data support the proposition that, in southern mid-autumn, the opacity was primarily dust that was well mixed in the lower few scale heights [Kahre et al., 2017]. In contrast, at other times of the year, the lowest part of the atmosphere was depleted in aerosols relative to higher altitudes.

Over $L_S$ 150-360° and 0-20°, the low-altitude depletion was only sparsely sampled and likely due to high-altitude dust transported over the crater following some regional storms. Using line-of-sight extinction measurements, Moore et al. [2019] found the crater to be somewhat dust depleted at most times of the year, especially around the onset of regional storms, but enhanced in dust during early southern summer. Smith et al. [2020] also found that the in-crater extinction did not respond quickly to increases in column opacity from regional storms, even when dust lifting in the crater contributed to opacity. Local dust lifting and the growth of the PBL above the crater rim [Fonseca et al., 2018] likely drove the low-altitude dust enhancement seen in line-of-sight extinction, while dust transported from storms elsewhere drove temporary high-altitude enhancements.

Over at least L$_S$ 90-135°, the lower relative abundance of aerosols at low altitudes was likely due to cloud and ice-haze layers at high altitudes, given the presence of the aphelion cloud belt (ACB) over the area (cf., Wolff et al., [2019]). In their Fig. 10, Smith et al. [2020] show that column optical depth near sols 1880 and 2550 (where the scale-height analysis, their study, and L$_S$ 90° overlap) was about one-third higher than would be predicted by line-of-sight opacity if the atmosphere were well mixed, which would be consistent with about 25% of the column optical depth typically coming from water ice.

### 4.5. Using near-Sun images for a halo search

Sky images associated with solar images were visually inspected for signs of ice, such as discrete clouds or halo-type features. With the blue or clear (RGB) filters, 537 M34 images were taken over sols 489-3638 that included the 22° scattering angle of the water-ice halo seen by the *Perseverance* rover [Lemmon et al., 2022a]. Discrete clouds appeared in several images during the L$_S$ 45-150° period in which the ACB can affect Gale crater [Campbell et al., 2020] but were not commonly seen in unprocessed images. The residual images showed more detail that we associated with ice hazes; however, these single, near-Sun images were not as effective in identifying and characterizing clouds as the time-lapse 'movies' discussed by Kloos et al. [2016; 2018] and Campbell et al. [2020]. No halo was seen in radiance-calibrated images or residual images after the removal of large-scale brightness gradients. As halos would indicate large, crystalline particles [Lemmon et al., 2022a], their absence suggests the particles are small (<11 μm) or amorphous, which is consistent with the ice types proposed by Clancy et al. [2003]. It is not apparent why haloes were not seen at Gale, but the southerly location near the margin of the ACB may result in insufficient water for the high supersaturations needed for rapid particle growth, or they may simply be rare events.

### 4.6. Sand flux

The cameras acted as aeolian sand collectors owing to a baffle design that effectively trapped sand. The cameras first showed signs of sand accumulation on sol 1576, and the maximum effects were seen over sols 1629 to 1634. A review of past images showed sand piles on the rover deck during an unusually low elevation (-81°) acquisition on sol 1197 that may indicate material was accumulated and dumped prior to recognition in the Tau data. The M-34 has a ~4.4 ×4.4 cm opening, while the M-100 opening is ~2.5×2.5 cm (for areas of 18.6 and 6.5 cm$^2$). In addition to the main opening, there are small holes at the top and bottom (which permitted access for screwdrivers during the attachment of the cameras to the RSM). Little material is likely to have entered these small holes. During operations and overnight storage, the average camera position was 20° below the horizon and within 90° of due south (toward the main dune field, see Supplementary Materials). The apparent apertures (corrected for the downward aim) were 17.7 and 6.0 cm$^2$, about 1.95 m above the surface. Based on baffle geometry, we estimate that a prismatic volume of approximately 2.6±0.4 cm$^3$ defined by the lip of the front aperture of the baffle and the gravity normal plane into the baffle defined by the 20° downward pointing is the likely maximum volume of aeolian sediment.  An independent estimate of the volume of the material was made by comparing a MAHLI image of the inside of the M-100 optics baffle with a view of that same area prior to launch (Figure 10). The area darkened by fines, adjusted for the oblique viewing angle, was ~19.5 cm$^2$, and the covering of steplike structures within the baffle implies a thickness, if uniform, of 0.2 cm at those steps, resulting in a volume of 3.9±1.5 cm$^3$. More likely, the thickness was nonuniform, and sand was at the base of obstacles. Assuming a 1.6±0.2 g cm$^{-3}$ for loose, unconsolidated sand, the M-100 held roughly 5.3±2.5 g, but the uncertainty is probably much larger than statistical.

Sand accumulation into the M-100 camera baffle is a direct indication of sand saltation across the rover traverse, but estimates of sand flux during strong wind events cannot be well constrained. Sand accumulation probably happened throughout the mission. Still, much of the observed volume was likely collected in about 60 to 160 sols, based on the rapid increase after it became significant on sol 1570 and the rover's position in an interdune area with aeolian activity after sol 1470. Thus, the cumulative sand collection rate would have been of order 0.3±0.2 kg m$^{-2}$ sol$^{-1}$. Unfortunately, wind speeds could not be measured, so the frequency of sand collection intervals is unknown, as are the wind strengths and durations of these intervals. The cumulative sand collection rate was 3±2 µg s$^{-1}$, but this does not include unknown effects of time spent with the cameras potentially facing away from sand-driving wind events, nor the complications from low-elevation camera observations that might have temporarily emptied part of the accumulated sand from the baffle.

Phase 2 of the Bagnold Dunes Campaign (BDC-2) occurred over sols 1603 to 1659 [Baker et al., 2018]. While earlier studies failed to show sediment motion, change detection images of the surface during BDC-2 showed daily migration of sand ripples comprising 100-µm diameter grains, generally toward the southwest [Baker et al., 2018]. Over the mission, most observed sand motion occurred overnight during southern summer in several km- to regional-scale winds [Baker et al., 2022]. The change-detection observations support the previous assumption that the sand accumulation happened in a short interval, as is often the case terrestrially, given that large sand fluxes require both a source of sand and the high winds of southern summer. In addition, around sol 1580, the Rover Environmental Monitoring Station ultraviolet sensor had the least dust cover of the mission after the first four months [Vicente-Retortillo et al., 2020], possibly owing to dust removal by sand scouring.

The cumulative sand collection rate for the M-100 baffle does not represent a total saltation flux, because collection occurred at a single height (~2 m). On Mars, trajectories of 100-µm grains in pure saltation are expected to reach heights below 1 m at wind friction speed, u*, of 1 m s$^{-1}$, even over rocky surfaces. However, these grains are small enough to be affected by turbulence and become partly suspended, enabling greater trajectory heights [Sullivan and Kok, 2017]. Larger grains saltating over hard surfaces are less subject to short-term suspension but achieve saltation heights >2 m due to upward grain momentum being retarded less efficiently by drag because grain mass grows faster than cross-sectional area as grain radius increases. Supplementary Figure S8 summarizes numerical simulations predicting that 150-, 200-, and 250-µm grains driven by $u* = 1$ m s$^{-1}$ should achieve saltation heights exceeding the ~2 m elevation of the M-100 camera baffle. Field studies where aeolian sediments have been collected at heights from a few cm to about 2 m show definite functional relations that permit estimating total flux from a single height Malin, [1984, 1985, 1991]. These field studies show that terrestrial saltation can reach several-meter heights over hard surfaces or surfaces with surface elements larger than pebbles [Sharp 1964, 1980; Malin and Eppler, 1981; Malin, 1984, 1985, 1991]. In Iceland, at a sand-poor, gravel-covered surface, the flux at ~140 cm height was 0.02 µg s$^{-1}$, and at a sand-rich, cobble- and boulder-covered surface, the flux at ~140 cm height was 12.7 µg s$^{-1}$ over a 24-day collection period. At the same sites, the flux at the same heights was 0.06 µg s$^{-1}$ and 4.2 µg s$^{-1}$ over a 1460-day collection period. Similar results have been seen in the Antarctic Dry Valleys and Trans-Antarctic Mountains. These field studies also reveal that, although the total flux decreases with height, the relative fraction of the coarser material increases with height. Laboratory measurements confirm these field observations, with the interpretation that the high trajectories result from considerable variation in surface normals [Liu et al., 2021]. The rover was on a rocky surface, with the surface including significant fractions of both bedrock and individual stones. These studies support the contention that the material in the baffle may comprise larger particles (sand or coarse sand) that are less influenced by atmospheric drag compared to the 100- to 150-µm median grain size seen by Baker et al. [2018] and Weitz et al. [2018] or the 100-µm grains modeled by Sullivan and Kok [2017]. New numerical simulations of saltation by one of us, using the framework of Sullivan and Kok [2017], showed that larger grains bounce to greater heights, with even 150-µm grains able to reach the baffle without bouncing off rover hardware and even 100-µm grains could reach relevant heights in short-term suspension. The grain size of the sand trapped inside the camera baffles could not be determined directly, but hand lens-quality images of sand accumulations on the rover deck help constrain the most abundant grain sizes saltating above that height and are consistent with the 100- to 150-µm grains seen in the ripples. The arm-mounted MAars Hand Lens Imager (MAHLI [Edgett et al., 2012]) acquired ~52 µm/pixel images of a partly sand-filled tray on the rover deck; the overwhelming majority of grains spanned 4 pixels or less, so were ≤200 µm (see Supplementary Information S9).

Alternatively, the winds that induced some saltation to 2-m heights may be different from the winds that caused most ripple migration. Daytime convective vortices commonly generated the highest winds seen by the rover, with the rover having a vortex encounter every 1-2 hours from 09:00 to 16:00 around L$_S$ 300° [Newman et al. 2019]. Vortices can lift sand from the surface at a rate of 10$^{-4}$ to >10$^{-1}$ kg m$^{-2}$ s$^{-1}$ [Neakrase et al., 2010]. Vortical winds would not systematically drive sand in a uniform direction, so the typical southward aim of the camera and southwestern ripple migration would not be problematic. Vortices were near a minimum during BDC-1, with <1 each sol and very few strong (>1 Pa) vortices compared to ~1.5 such vortices per sol during BDC-2.

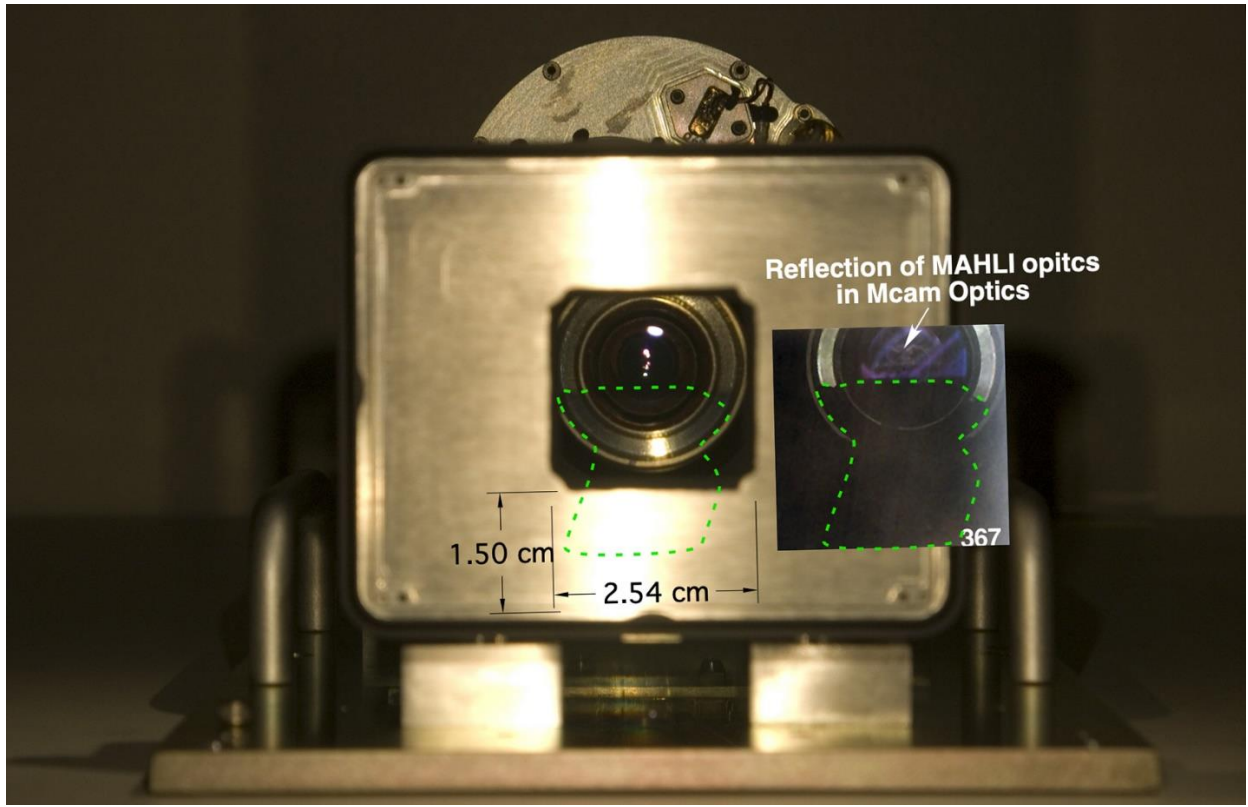

Figure 10. View from front of M-100 in the laboratory, with inset from MAHLI image 1749MH0007200210700397C00. Green dashes show the extent of a dark area interpreted to be aeolian fines inside the M-100 sunshade. The Mastcam elevation was approximately +40°.

## 5. Conclusions

The *Curiosity* rover's Mastcam has monitored atmospheric optical depth at two wavelengths, 440 and 880 nm, for over five Mars years. The 880-nm measurement was mostly a dust optical depth record, although the dust-ice ratio was not determined. Observed optical depth trends were consistent with those seen by other surface and orbital assets at Mars. There were decreasing optical depths over $L_S$ 0-140°, with low sol-to-sol or year-to-year variability; elevated optical depths with occasional significant increases due to dust storms over $L_S$ 140-210°; and further increases in dust near $L_S$ 210° and 315°, followed by slow decays. During the second half of the year, diurnal optical depth variations were consistent with the 10% pressure variations observed in the crater. There was little evidence for consistent daily trends during the first part of the year, except for the period after one early dust storm around $L_S$ 150° of MY 36. The color ratio of the optical depth was mostly constant throughout the year, except for a repeated but noisy indication of relatively lower blue optical depths around $L_S$ 90-130°, around the peak of the ACB season. Low-elevation solar images did not indicate a consistently well-mixed atmosphere, requiring high-altitude dust enhancement during storms and suggesting that ~25% of the column optical depth was high-altitude ice during the ACB season. While ice hazes and clouds were seen in near-Sun sky images, we found no evidence of scattering halos. Near sol 1600, enough sand had collected in the 1.97-m high sunshades that the cameras could no longer aim toward the zenith without obscuration. The imaging strategy was adapted to avoid such geometries.

# Acknowledgments

We are grateful to the teams that developed, landed, and operated *Curiosity* on Mars, allowing for the present study. The research was conducted partly at the Jet Propulsion Laboratory, California Institute of Technology, under a contract with the National Aeronautics and Space Administration (80NM0018D0004). MTL was supported via sub-contract 18-1187 from Malin Space Science Systems, Inc. SDG was supported by the MSL Participating Scientist program. JMB was supported by MSL Participating Scientist Grant 80NSSC22K0657. AV-R was supported by the Comunidad de Madrid Project S2018/NMT-4291 (TEC2SPACE-CM). M-PZ was supported by grant PID2019-104205GB-C21 funded by MCIN/AEI/10.13039/501100011033. JM-T was supported by UK Space Agency projects ST/W00190X/1 and ST/V00610X/1.